# Potential–driven adiabatic connection in density functional theory

Andreas Savin

Laboratoire de Chimie Théorique, UMR 7616
CNRS and Université Pierre et Marie Curie – Paris VI
4, place Jussieu
F–75252 Paris Cedex 5, France

∎ **Abstract**

As density functional theory conventionally assumes that the density of a chosen model system (e.g., the Kohn–Sham system) is the same as the exact one, one might expect that approximations to the exact density introduce supplementary errors by falsifying the density. In fact, this is not true: by modelling the exchange–correlation holes for all densities, density functional approximations avoid this problem. The technique used to show it is a potential–driven adiabatic connection which hopefully will also permit constructing new approximations in the spirit of DFT.

∎ **Introduction**

### DFT

In density functional theory (DFT) the Schrödinger equation is solved for model systems where the interaction between particles is fictitious, i.e., not the physical, Coulomb one . In the Kohn–Sham model [1], for example, the interaction is reduced to its simplest form: it is set to zero. The energy of the model is, of course, different from that of the physical system. Insight into the nature of this difference, which is needed for obtaining the energy, can be obtained by considering an 'adiabatic connection', a process in which the interaction is progressively modified between that of the of the model to that of the physical system [2–5]. The evolution between the model and the physical system can be characterized by a parameter, $\lambda$ which varies between $\lambda_0$, characterizing the model, and $\lambda_1$ characterizing the physical system.

In its most widespread formulation, the expression of the correction to the model energy, needed to obtain the exact energy, contains the evolution of the pair density, $P_2(r_1, r_2)$ along the adiabatic connection. Modelling $P_2(r_1, r_2)$ was not only success–fully used for constructing many of the density functional approximations (see,e.g., Refs. [6–12]), but is also explicitly used in methods like the random phase approximation (see, e.g., Refs. [13, 14]). In its most widespread variant, although $P_2$ is a function of $\lambda$, the one–particle density $n(r)$ does not vary along the adiabatic connection [3].

**Model densities may not be exact**

With a density functional approximation (DFA) the model does not yield the exact density; in general, $n^{\lambda_0} \neq n^{\lambda_1}$. In the following, a simple example will be given. It can be considered exaggerated, but it has the advantage that accurate numbers are known for it [15]. The system of two non–interacting particles in the potential $-\zeta/r$ will be assumed to be an approximation of the exact Kohn–Sham system which yields the density of the He atom. The choice $\zeta = 1.344..$ yields the exact asymptotic decay of the density of the He atom. This model system thus reproduces exactly only a given property of the physical density, not the physical density itself. One can numerically construct a system in which the interaction between electrons is of Coulomb type, but has the density of this model, $n^{\lambda_0} = 2\left(\zeta^3/\pi\right)\exp(-\zeta r)$ [15] and calculate the ground state energy of the fully interacting system having this density, in the external potential of the He atom. Keeping the density constant produces a very large error in the total energy ($\approx 0.13$ hartree). Not surprisingly, the largest error comes from the electrostatic part of the electron–electron interaction, $\approx 0.37$ hartree and the difference in the one–particle part of the energy is of the same order of magnitude, $\approx 0.3$ hartree. If one concentrates, however, on the parts which are modelled in DFT, the situation is better: the correlation energy obtained for the physically interacting system with $n^{\lambda_0}$ is $-0.043$ hartree, reasonably close to that of the He atom (see, e.g., Ref. [16]), $-0.042$ hartree. However, the exchange energy differs considerably; it is $-0.840$ hartree [15] for the system with $n^{\lambda_0}$, vs. $-1.025$ for the He atom (see, e.g., Ref. [16]).

This example seems to support the idea that one should take into account the change of the density between the model and the physical system.

**Objective**

The objective of this paper is to show that, with a slightly modified adiabatic connection, the problem of variable density is in fact avoided by DFAs. The key idea is that one has to take into account that DFAs provide models for all densities. Furthermore, it will be argued that the modified adiabatic connection allows going beyond DFAs in a systematic way. Of course, like in all methods of Quantum Chemistry, this last step has to be payed with more computational effort.

## ■ The modified adiabatic connection

**Family of Hamiltonians**

Let us consider a family of model Hamiltonians, $\lambda_0 \leq \lambda \leq \lambda_1$

$$H^\lambda \equiv H(v^\lambda, w^\lambda) = T + V^\lambda + W^\lambda \tag{1}$$

where

$$V^\lambda = \sum_{i=1,N} v^\lambda(r_i) \tag{2}$$

is a local one–particle potential,

$$V_{ne} = \sum_{i=1,N} v_{ne}(r_i) \tag{3}$$

is the physical local one–particle potential $\left(v^{\lambda_1} \equiv v_{ne}\right)$,

$$W^\lambda = \sum_{i<j} w(r_i, r_j) \tag{4}$$

is an operator which describes a fictitious two-particle interaction, which becomes, for $\lambda = \lambda_1$, the physical two–particle interaction

$$V_{ee} = \sum_{i<j} v_{ee}(r_i, r_j) \tag{5}$$

$v_{ee}(r, r') = 1/|r - r'|$, in hartree atomic units.

In order to compare with DFT, $v^\lambda$ is further decomposed:

$$v^\lambda(r) = v_{ne}(r) + v_h\bigl(r; n, v_{ee} - w^\lambda\bigr) + v_{xc}{}^\lambda(r) \tag{6}$$

where

$$v_h(r; n, w) \equiv \int d^3 r'\, n(r')\, w(r, r') \tag{7}$$

and $v_{xc}{}^\lambda$ is defined by Eq. 6. Below, for analyzing approximations, we will proceed in a different way: we will choose some $v_{xc}{}^\lambda(r)$, and use Eq. 6 to define $v^\lambda$; $v_{xc}{}^\lambda$ will be chosen to vanish as $\lambda \to \lambda_1$, to ensure that $v^{\lambda_1} \to v_{ne}$.

**Energy expression**

To obtain the total energy we will write, also in analogy to DFT,

$$E = \langle \Psi^\lambda | T + V_{ne} + W^\lambda | \Psi^\lambda \rangle + U[n^\lambda; v_{ee} - w^\lambda] + E_{xc}[v^\lambda, w^\lambda] \tag{8}$$

where

$\Psi$ is an antisymmetric wave function; $\Psi^\lambda$ will be used as a notation for a minimizing $\Psi$. $\Psi^\lambda$ yields the one–particle density $n^\lambda(r)$. Furthermore,

$$U[n; w] \equiv \frac{1}{2} \int\!\!\int n(r)\, n(r')\, w(r, r') \tag{9}$$

is a Hartree (i.e., electrostatic) term and $E_{xc}$ is defined by Eq. 8.

Please notice that the last two terms on the r.h.s. of Eq. 8 vanish when $\lambda = \lambda_1$.

**Variation with $\lambda$**

In order to study the change w.r.t. $\lambda$, after taking the derivative of Eq. 8 w.r.t. $\lambda$, we get

$$0 = \langle \Psi^\lambda | \partial_\lambda W^\lambda | \Psi^\lambda \rangle - U[n^\lambda; \partial_\lambda w^\lambda] - \int v_{xc}{}^\lambda\, \partial_\lambda n^\lambda + \partial_\lambda E_{xc}[v^\lambda, w^\lambda] \tag{10}$$

To obtain Eq. 10, have used the variational character of $\Psi^\lambda$ for $H^\lambda$,

$$\partial_\lambda \langle \Psi^\lambda | T + V_{ne} + W^\lambda | \Psi^\lambda \rangle = \partial_\lambda \langle \Psi^\lambda | T + V^\lambda + W^\lambda | \Psi^\lambda \rangle + \partial_\lambda \int d^3 r\, (v_{ne} - v^\lambda)\, n^\lambda =$$
$$\langle \Psi^\lambda | \partial_\lambda V^\lambda + \partial_\lambda W^\lambda | \Psi^\lambda \rangle - \int d^3 r\, n^\lambda\, \partial_\lambda v^\lambda + \int d^3 r\, (v_{ne} - v^\lambda)\, \partial_\lambda n^\lambda = \tag{11}$$
$$\langle \Psi^\lambda | \partial_\lambda W^\lambda | \Psi^\lambda \rangle + \int d^3 r\, (v_{ne} - v^\lambda)\, \partial_\lambda n^\lambda,$$

and

$$\partial_\lambda U[n^\lambda; v_{ee} - w^\lambda] = \int d^3 r\, v_h(r; n^\lambda, v_{ee} - w)\, \partial_\lambda n^\lambda - \frac{1}{2} \int\!\!\int d^3 r\, d^3 r'\, n(r)\, n(r')\, \partial_\lambda w^\lambda, \tag{12}$$

as well as Eq. 6.

### Integrated formulas

Eq. 10 can be integrated over $\lambda$, between $\lambda_0$ and $\lambda_1$, to yield

$$E_{xc}[v^{\lambda_0}, w^{\lambda_0}] = \int_{\lambda_0}^{\lambda_1} d\lambda \left( \langle \Psi^\lambda | \partial_\lambda W^\lambda | \Psi^\lambda \rangle - U[n^\lambda; \partial_\lambda w^\lambda] - \int v_{xc}(r; n^\lambda, v_{ee} - v^\lambda) \partial_\lambda n^\lambda \right) \quad (13)$$

One can also use the exchange–correlation part of the pair density produced by $\Psi^\lambda$, $P_2(r_1, r_2; \Psi^\lambda)$,

$$P_{xc}(r_1, r_2; \Psi^\lambda) = P_2(r_1, r_1; \Psi^\lambda) - n^\lambda(r_1) n^\lambda(r_2) \quad (14)$$

to re–write Eq. 13 as

$$E_{xc}[v^{\lambda_0}, w^{\lambda_0}] = \int_{\lambda_0}^{\lambda_1} d\lambda \frac{1}{2} \iint d^3 r \, d^3 r' \, P_{xc}(r_1, r_2; \Psi^\lambda) \partial_\lambda w^\lambda(r, r') - \int_{\lambda_0}^{\lambda_1} d\lambda \int v_{xc}^\lambda(r) \partial_\lambda n^\lambda(r) \quad (15)$$

### Hamiltonian–driven adiabatic connections

In the derivation of the formulas above, the adiabatic connection was driven by the change in the Hamiltonian, in particular by the change of the one– and two–body potential: it was potential–driven. One can also produce model Hamiltonians by changing the kinetic energy operator. For example, one can keep the two–body operator equal to $V_{ee}$ for all $\lambda$ (cf., e.g., Ref. [17]). With such a one–body–operator driven adiabatic connection one can produce expressions for the correlation energy which depend on the one–body density matrix along the adiabatic connection, $\gamma(r, r'; \Psi^\lambda)$.

This type of adiabatic connection will not be discussed here, as it is only seldom used to produce approximations to the universal correlation energy functional.

## ■ Relationship to DFT

### Relationship to exact DFT

The adiabatic connection formula of DFT is well–known [3]. The resulting equation has the same form as Eq. 15 without the last term on the r.h.s., as $n$ does not vary with $\lambda$. One should keep in mind that in DFT $E_{xc}$ is not determined by $v^{\lambda_0}$ and $w^{\lambda_0}$, but by $n$ and $w^{\lambda_0}$,

$$E_{xc}[n, w^{\lambda_0}] = \int_{\lambda_0}^{\lambda_1} d\lambda \frac{1}{2} \iint d^3 r \, d^3 r' \, P_{xc}(r, r'; \Psi^\lambda) \partial_\lambda w^\lambda(r, r') \quad (16)$$

and that $v^\lambda$ is constructed by using in Eq. 6

$$v_{xc}^\lambda(r) = \delta E_{xc}[n, w^{\lambda_0}] / \delta n(r). \quad (17)$$

### Density functional approximations

In practice, DFAs are made to define the model systems: $E_{xc}$ is in general replaced by some approximation, $\tilde{E}_{xc}$:

$$\tilde{E}_{xc}[\tilde{n}^{\lambda_0}, \lambda_0] = \int_{\lambda_0}^{\lambda_1} d\lambda \frac{1}{2} \iint d^3 r \, d^3 r' \, \tilde{P}_{xc}(r, r'; \tilde{n}^{\lambda_0}(r)) \partial_\lambda w^\lambda(r, r') \quad (18)$$

$\tilde{P}_{xc}$ is some model for $P_{xc}$; in LDA, for example, it is that of the uniform electron gas with density $\tilde{n}^{\lambda_0}(r)$; $\tilde{n}^{\lambda_0}$ is the density obtained from $\tilde{\Psi}^{\lambda_0}$ which in turn depends on $w^{\lambda_0}$ and $v^{\lambda_0}$. For the latter, Eq. 6 is used, and

$$\tilde{v}_{\text{xc}}{}^{\lambda_0}(r; n, w) \equiv \delta \tilde{E}_{\text{xc}}[n, \lambda_0]/\delta n(r) \tag{19}$$

at $n = n^{\lambda_0}$.

Comparing the equation of DFT, Eq. 16, with that of DFAs, Eq. 18, one can notice two differences: i) the latter uses of a model, and ii) the model uses $n^{\lambda_0}$ instead of $n^\lambda$. The latter point arises as the DFT assumption was made in the derivation, namely that $n$ was assumed not to change with $\lambda$. (When a calculation at $\lambda_0$ is done, only $n^{\lambda_0}$ is known, so that the information about $n^{\lambda(>\lambda_0)}$ is missing.)

$\partial_\lambda \tilde{E}_{\text{xc}}[n^\lambda; \lambda]$

Notice that $\tilde{E}_{\text{xc}}$ depends on $\lambda$ implicitly, via the $\lambda$–dependence of $n^\lambda$, and explicitly, for a given $\lambda$, as the functional will change for the same density, as the interaction $w^\lambda$ changes with $\lambda$:

$$\partial_\lambda \tilde{E}_{\text{xc}}[n^\lambda, \lambda] = \partial_\lambda \tilde{E}_{\text{xc}}[n^\lambda, \tilde{\lambda}]\big|_{\tilde{\lambda}=\lambda} + \partial_\lambda \tilde{E}_{\text{xc}}[n, \lambda]\big|_{n=n^\lambda} \tag{20}$$

The first term on the r.h.s. is the derivative of $\tilde{E}_{\text{xc}}$ at fixed $\lambda$ which by the chain rule is

$$\partial_\lambda \tilde{E}_{\text{xc}}[n^\lambda, \tilde{\lambda}]\big|_{\tilde{\lambda}=\lambda} = \int \tilde{v}_{\text{xc}}(r; n^\lambda, w^\lambda)\, \partial_\lambda n^\lambda(r) \tag{21}$$

while the last term on the r.h.s. of Eq. 20, the derivative at fixed $n$, is , by using Eq. 18,

$$\partial_\lambda \tilde{E}_{\text{xc}}[n, \lambda]\big|_{n=n^\lambda} = -\frac{1}{2} \iint d^3 r\, d^3 r'\, \tilde{P}_{\text{xc}}(r, r'; n^\lambda(r))\, \partial_\lambda w^\lambda(r, r') \tag{22}$$

**Special choice for $v_{\text{xc}}{}^\lambda$**

We now go back to the adiabatic connection in which the density is allowed to vary, but specify now a potential which until now was arbitrary. We choose

$$v_{\text{xc}}{}^\lambda(r) = \tilde{v}_{\text{xc}}(r; n^\lambda, w^\lambda) \tag{23}$$

By this choice,

$$\begin{aligned} \Psi^\lambda &= \tilde{\Psi}^\lambda \\ n^\lambda &= \tilde{n}^\lambda \end{aligned} \tag{24}$$

**Adiabatic connection for the special choice of $v_{\text{xc}}{}^\lambda$**

We can use Eq. 23 in Eq. 15, next use Eq. 21, followed by Eq. 20, and finally use Eq. 22

$$E_{xc}[v^{\lambda_0}, w^{\lambda_0}] = \int_{\lambda_0}^{\lambda_1} d\lambda \frac{1}{2} \int\int d^3r\, d^3r'\, P_{xc}(r_1, r_2; \Psi^\lambda) \partial_\lambda w^\lambda(r, r') - \int_{\lambda_0}^{\lambda_1} d\lambda \int \tilde{v}_{xc}(r; n^\lambda, w^\lambda) \partial_\lambda n^\lambda(r) =$$

$$\int_{\lambda_0}^{\lambda_1} d\lambda \frac{1}{2} \int\int d^3r\, d^3r'\, P_{xc}(r_1, r_2; \Psi^\lambda) \partial_\lambda w^\lambda(r, r') - \int_{\lambda_0}^{\lambda_1} d\lambda (\partial_\lambda \tilde{E}_{xc}[n^\lambda, \tilde{\lambda}]|_{\tilde{\lambda}=\lambda}) =$$

$$\int_{\lambda_0}^{\lambda_1} d\lambda \frac{1}{2} \int\int d^3r\, d^3r'\, P_{xc}(r_1, r_2; \Psi^\lambda) \partial_\lambda w^\lambda(r, r') - \int_{\lambda_0}^{\lambda_1} d\lambda (\partial_\lambda \tilde{E}_{xc}[n^\lambda, \lambda] - \partial_\lambda \tilde{E}_{xc}[n, \lambda]|_{n=n^\lambda}) =$$

$$\int_{\lambda_0}^{\lambda_1} d\lambda \frac{1}{2} \int\int d^3r\, d^3r'\, P_{xc}(r_1, r_2; \Psi^\lambda) \partial_\lambda w^\lambda(r, r') -$$

$$\int_{\lambda_0}^{\lambda_1} d\lambda (\partial_\lambda \tilde{E}_{xc}[n^\lambda, \lambda]) - \int_{\lambda_0}^{\lambda_1} d\lambda \frac{1}{2} \int\int d^3r\, d^3r'\, \tilde{P}_{xc}(r, r'; n^\lambda(r)) \partial_\lambda w^\lambda(r, r')$$

so that, with $\tilde{E}_{xc}[n^{\lambda_0}, \lambda_1] = 0$,

$$E_{xc}[v^{\lambda_0}, w^{\lambda_0}] - \tilde{E}_{xc}[n^{\lambda_0}, \lambda_0] = \int_{\lambda_0}^{\lambda_1} d\lambda \frac{1}{2} \int\int d^3r\, d^3r'\, (P_{xc}(r_1, r_2; \Psi^\lambda) - \tilde{P}_{xc}(r, r'; n^\lambda(r))) \partial_\lambda w^\lambda(r, r') \quad (26)$$

**Interpretation**

Eq. 26 shows that the error of the model is due the difference between $P_{xc}(\Psi^\lambda)$ and the model $\tilde{P}_{xc}(n^\lambda)$. As in DFA the models are defined to work for all densities (as one does not know beforehand what density is of interest) they also work for $n^\lambda$. Thus, from the perspective of Eq. 26 (and of the modified adiabatic connection) there is no need for any supplementary correction due to density changes.

**Relevance for DFAs**

When the ground state energy is computed, Eq. 26 tells us that we can comfortably ignore the fact that the density of the model is not the exact one — as long as the DFA is based upon a hole model. Most existing approximations used (the local density approximation, LDA, most of the generalized gradient approximations, GGAs, etc.) are based on hole models.

Notice also that in 'density functional calculations' sometimes potentials are used which are not derivatives of a functional of the density. They show up, e.g., when optimized effective potentials are used, or when making approximations for time–depe–nent DFT, e.g., for correcting the asymptotic behavior of the approximate Kohn–Sham potential. In such situation one leaves the standard frame of DFT, but not that of the present approach.

## ∎ Perspectives

**Losses and gains**

A Hamiltonian–driven adiabatic connection is identical to the adiabatic connection in DFT when the density is kept constant. When the density is not kept fixed, it looses the pure beauty of DFT. However, the added flexibility might not only bring closer theory to what is done in practice in DFAs, but also might give some hints about how to improve approximations. Finally, many of the successful DFAs were constructed from hole models, and they can continue to be used in the potential–driven adiabatic connection.

**Choosing $v^\lambda$**

Analyzing $v^\lambda$ was used over the years to understand DFAs which are normally constructed by using an ansatz of the form

$$\tilde{E}_{xc}[n, \lambda_0] = \int d^3 r\, n(r)\, \partial_\lambda \tilde{\varepsilon}_{xc}^\lambda(n(r),\ |\nabla n(r)|^2,\ ...) \tag{27}$$

Unfortunately, the equality does not suffice to define $\tilde{\varepsilon}_{xc}$: the l.h.s. is a number, while $\tilde{\varepsilon}_{xc}$ is a function. (In other words, any function which multiplied with $n$ integrates to zero can be added to $\tilde{\varepsilon}_{xc}$ without changing the value of the integral.) However, in DFT, one can compare safely, for a given system, the accurate $v_{xc}^\lambda$, Eq. 17, with that obtained from approximations. Thus, one can also use the knowledge gained in the last years for constructing accurate $v_{xc}^\lambda$ in DFT for constructing $v^\lambda$ for the potential–driven adiabatic connection.

By the requirement of using model systems having as ground state density the exact one, DFT ensures that the model system is in most cases sufficiently close to the exact one, e.g., has the exact electrostatic energy. In practice, however, as the exact density is unknown, DFAs produce only 'reasonable' densities. Thus, to have similar performance in the potential–driven adiabatic connection, the $v^\lambda$ should yield 'reasonable' densities, and thus be sufficiently 'close' to the $v^\lambda$ which keep the density constant. As the terms 'reasonable' and 'close' are not well–defined, the choice of $v^\lambda$ is left to further exploration. It is possible to perform calculations in the spirit of DFT without the constraint of using potentials which are derivatives of some density functional. To start the explorations, however, one can imagine using forms of $v^\lambda$ similar to those existing in DFAs; a few parameters in $v^\lambda$ could be determined 'on the flight', i.e., made system specific, e.g., by using perturbation theory, see below.

**State following**

In the potential–driven adiabatic connection the model system does not have to be in its ground state. (Of course, the model for the pair density will have to show some dependence on the state chosen, e.g., by a dependence on the depth of the exchange–correlation hole, cf. Refs. [18–20].) By convenient choices of $v^\lambda$, it should not only be possible to follow a given state along the adiabatic connection, but also to avoid some of the surprises produced in model systems keeping the ground state density constant (the change of the nature of the ground state, artificial degeneracies, missing degeneracies, jumps, ...). As the potential–driven adiabatic connection has more flexibility, size–consistency problems as those presented in [21], might also be avoided.

**Perturbation series for improving $E_{xc}^\lambda$**

The idea to use perturbation theory to improved density functionals (see, e.g. [22]) can also borrowed for the present context. Considering the system at $\lambda_0$, defined by the Hamiltonian $H^{\lambda_0}$, one can recover information about the system at $H^\lambda$ by using perturbation series where the perturbation operator is $H^\lambda - H^{\lambda_0}$. To obtain the first–order correction to the energy might be not very expensive, as only the wave function (or the reduced density matrices) at $\lambda_0$ are needed. The 'slope of the correction' being now known, can be used for improving 'on the flight' $E_{xc}$, e.g., by re–adjusting the depth of the exchange–correlation hole.

Such an approach to correct $E_{xc}$ was already used with a different adiabatic connection, where the one–electron part of the Hamiltonian is modified, $T^\lambda + V^\lambda + V_{ee}$. [17]

By a change of viewpoint one can see the perturbation series as resulting from taking derivatives w.r.t. $\lambda$. Taking the first derivative w.r.t. $\lambda$ in Eq. 8 is equivalent to considering the first–order perturbation term. Higher derivatives are related to higher orders in perturbation theory. One can use, as when deriving Eq. 10, that the total physical energy, $E$, is independent of $\lambda$. This method yields further equations which can be used to constrain $E_{xc}$ using information specific to the system.

The adiabatic connection, Eq. 13 or Eq. 15, amounts to replace the perturbation expansion in one point with first–order perturbation corrections on all points between $\lambda_0$ and $\lambda_1$. This suggests that one could repeat the calculation at a new $\lambda_0$, say $\lambda_0'$, and use both the information at $\lambda_0$ and $\lambda_0'$ to improve an existing ansatz for an approximation of the exchange–correlation term. This could be useful, for instance, if we want either to avoid the effort of higher order perturbation calculations, or we don't trust the perturbation expansion to higher order.

Still another way to exploit the adiabatic connection is to use different potentials, and to compute different corrections to it, starting at the same $\lambda_0$. Of course, in principle, the physical energy, $E$, is independent of the path chosen. This constraint may be a path to further improve the approximations for $E_{xc}$.

**Acknowledgments**

Stimulating discussions with Paul Ayers (McMaster University, Hamilton, Canada) , Paola Gori–Giorgi (CNRS, Paris, France) and Julien Toulouse (Université Pierre et Marie Curie – Paris VI, Paris, France) are gratefully acknowledged. Financial support was granted by the ANR–07–BLAN–0272–03.

It is a pleasure to dedicate this paper to John Perdew who with his work, deep understanding, and infinite patience has helped so many of us to enjoy DFT.